\documentclass[12pt]{article}

\usepackage{epsfig}

\usepackage{amssymb, amsfonts}

\usepackage{verbatim}

\def\ex{{{\rm e}}}

\def\tr{{\rm tr}\,}

\def\e{{\mathbf e}}

\newtheorem{theorem}{Theorem}[section]

\newtheorem{examp}{Example}[section]

\newtheorem{coroll}{Corollary}[section]

\newtheorem{examps}{Examples}[section]

\newtheorem{lemma}{Lemma}[section]

\newtheorem{remark}{Remark}[section]

\newtheorem{remarks}[remark]{Remarks}

\newtheorem{proposition}{Proposition}[section]

\newtheorem{definition}{Definition}[section]

\def\br{\begin{remark}\rm\small}

\def\1{{\bf 1}}

\def\er{\end{remark}}

\def\bt{\begin{theorem}\rm}

\def\et{\end{theorem}}

\def\bc{\begin{coroll}\rm}

\def\ec{\end{coroll}}

\def\brs{\begin{remarks}.\\ \rm\small\begin{enumerate}}

\def\ers{\end{enumerate}\end{remarks}}

\def\bx{\begin{examp}\small}

\def\ex{\end{examp}}

\def\bl{\begin{lemma}\small}

\def\el{\end{lemma}}

\def\bxs{\begin{examps}. \rm\begin{enumerate}}

\def\exs{\end{enumerate}\end{examps}}

\def\bd{\begin{definition}}

\def\ed{\end{definition}}

\def\bp{\begin{proposition}\rm}

\def\ep{\end{proposition}}

\def\beq{\begin{equation}}

\def\eeq{\end{equation}}

\def\bea{\begin{eqnarray}}

\def\eea{\end{eqnarray}}

\def\beas{\begin{eqnarray*}}

\def\eeas{\end{eqnarray*}}

\def\pa{\partial}

\def\L{\Lambda}
\renewcommand\l{\lambda}

\def\Res{{\rm Res\,}}

\def\curve{{\cal E}}

\def\br{\begin{remark}\rm\small}
\def\er{\end{remark}}
\def\bt{\begin{theorem}}
\def\et{\end{theorem}}
\def\bd{\begin{definition}}
\def\ed{\end{definition}}
\def\bp{\begin{proposition}}
\def\ep{\end{proposition}}
\def\bl{\begin{lemma}}
\def\el{\end{lemma}}
\def\bc{\begin{corollary}}
\def\ec{\end{corollary}}
\def\beaq{\begin{eqnarray}}
\def\eeaq{\end{eqnarray}}

\newcommand{\R}{{\mathbb{R}}}
\newcommand{\C}{{\mathbb{C}}}
\newcommand{\Z}{{\mathbb{Z}}}

\newcommand{\virg}{\quad , \quad}

\newcommand{\om}{\omega}
\newcommand{\Om}{\Omega}
\newcommand{\calD}{{\cal D}}
\newcommand{\xbar}{\overline{x}}
\newcommand{\rhobar}{\overline{\rho}}
\newcommand{\pbar}{\overline{p}}

\begin{document}

\pagestyle{empty}
\hfill SPhT-T05/036
\addtolength{\baselineskip}{0.20\baselineskip}
\begin{center}
\vspace{26pt}
{\large \bf {Large $N$ asymptotics of orthogonal polynomials\\
From integrability to algebraic geometry}}
\newline
\vspace{26pt}

{\sl B.\ Eynard}\hspace*{0.05cm}\footnote{ E-mail:  }
\vspace{6pt}
Service de Physique Th\'{e}orique de Saclay,\\
F-91191 Gif-sur-Yvette Cedex, France.\\
\end{center}

%

%
%

%

%
%

%
%

%
\section{Introduction}

Random matrices play an important role in physics and mathematics \cite{Mehta, courseynard, BI, DGZ, Guhr, Moerbeke:2000, DeiftBook}.
It has been observed more and more in the recent years how deeply random matrices are related to integrability ($\tau$-functions),
and algebraic geometry.

Here, we consider the computation of large n asymptotics for orhogonal polynomials as an example of a problem where the concepts of integrability,
isomonodromy and algebraic geometry appear and combine.

The method presented here below, is not, to that date, rigorous mathematicaly. It is based on the asumption that an integral with a large number of variables
can be approximated by a  saddle-point method. This asumption was never proven rigorously, it is mostly based on ``physical intuition''.
However, the results given by that method have been rigorously proven by another method, namely the Riemann--Hilbert method \cite{BlIt, BlIt1, dkmvz, dkmvz2}.
The method presented below was presented in many works \cite{eynchain, eynchaint, BEHAMS, eynbetapol, eynhabilit}.

\section{Definitions}

Here we consider the 1-Hermitean matrix model with polynomial potential:

\bea\label{defZ}
Z_N&:=& \int_{H_N} dM\, \e^{-N\tr V(M)}\cr
&=& \int_{\R^N} dx_1\dots dx_N\,\, \left(\Delta(x_1,\dots,x_N)\right)^2 \,\, \prod_{i=1}^N \e^{-NV(x_i)}
\eea
where $\Delta(x_1,\dots,x_N):=\prod_{i>j} (x_i-x_j)$, and the $x_i$'s are the eigenvalues of the matrix $M$,
and $V(x)$ is a polynomial called the potential:
\beq\label{defV}
V(x) = \sum_{k=0}^{\deg V} g_k x^k
\eeq

\begin{remark}
All the calculations which are presented below, can be extended to a more general setting, with no big fundamental changes:

- one can consider $V'(x)$ any rational fraction \cite{BEHsemiclas} instead of polynomial, in particular one can add logarithmic terms to the potential $V(x)$.

- one can consider arbitrary paths (or homology class of paths) of integrations $\Gamma^N$ insteaf of $\R^N$,
in particular finite segments \cite{marcopath} ...

- one can study non hermitean matrix models \cite{eynbetapol}, where the Vandermonde $\Delta^2$ is replaced by $\Delta^\beta$ where $\beta=1,2,4$.

- one can consider multi-matrix models, in particular 2-matrix model \cite{BEHAMS, eynchain, eynchaint}.
\end{remark}

\section{Orthogonal polynomials}

Consider the family of monic polynomials $p_n(x)=x^n + O(x^{n-1})$, defined by the orthogonality relation:
\beq
\int_{\R} p_n(x) p_m(x) \e^{-NV(x)} dx = h_n \delta_{nm}
\eeq
It is well known that the partition function is given by \cite{Mehta}:
\beq
Z_N = N! \prod_{n=0}^{N-1} h_n
\eeq

Such an orthogonal family always exists if the integration path is $\R$ or a subset of $\R$, and if the potential is a real polynomial.
In the more general setting, the orthogonal polynomials ``nearly always'' exist (for arbitrary potentials, the set of paths for which they don't exist is enumerable).

\medskip

We define the kernel:
\beq
K(x,y):=\sum_{n=0}^{N-1} {p_n(x) p_n(y)\over h_n}
\eeq

One has the following usefull theorems:
\bt Dyson's theorem \cite{thDyson}:
any correlation function of eigenvalues, can be written in terms of the kernel $K$:
\beq
\rho(\l_1,\dots, \l_k) = \det(K(\l_i,\l_j))
\eeq
\et
Thus, if one knows the orthogonal polynomials, then one knows all the correlation functions.

\bt Christoffel-Darboux theorem \cite{Mehta, Szego}:
The kernel $K(x,y)$ can be written:
\beq
K(x,y) = \gamma_N\,{p_N(x)p_{N-1}(y)-p_N(y)p_{N-1}(x)\over x-y}
\eeq
\et
Thus, if one knows the polynomials $p_N$ and $p_{N-1}$, then one knows all the correlation functions.

\medskip

Our goal now, is to find large $N$ ''strong'' asymptotics for $p_N$ and $p_{N-1}$, in order to have the large $N$ behaviours of any correlation functions.

\medskip
{\bf Notation:}
we define the wave functions:
\beq
\psi_n(x) := {1\over \sqrt{h_n}}\, p_n(x)\, \e^{-{N\over 2}V(x)}
\eeq
they are orthonormal:
\beq
\int \psi_n(x)\psi_m(x) = \delta_{nm}
\eeq

\section{Differential equations and integrability}

It can be proven that $(\psi_n,\psi_{n-1})$ obey a differential equation of the form \cite{bonan, BlIt, Mehta, TW2, BEHtauiso}:
\beq
-{1\over N} \,{\pa \over \pa x} \pmatrix{\psi_n(x) \cr \psi_{n-1}(x)} = {\calD}_n(x)\,\pmatrix{\psi_n(x) \cr \psi_{n-1}(x)}
\eeq
where $\calD_n(x)$ is a $2\times 2$ matrix, whose coefficients are polynomial in $x$, of degree at most $\deg V'$.
(In case $V'$ is a rational function, then $\calD$ is a rational function with the same poles).

\medskip
$(\psi_n,\psi_{n-1})$ also obeys differential equations with respect to the parameters of the model \cite{BlIt, BEHtauiso}, i.e. the coupling constants, i.e. the
$g_k$'s defined in \ref{defV}:
\beq
{1\over N} \,{\pa \over \pa g_k} \pmatrix{\psi_n(x) \cr \psi_{n-1}(x)} ={\cal U}_{n,k}(x)\,\pmatrix{\psi_n(x) \cr \psi_{n-1}(x)}
\eeq
where ${\cal U}_{n,k}(x)$ is a $2\times 2$ matrix, whose coefficients are polynomial in $x$, of degree at most $k$.

\medskip
It is also possible to find some discrete recursion relation in $n$ (see \cite{BEHtauiso}).

\medskip
The compatibility of these differential systems, i.e. ${\pa\over \pa x}{\pa\over \pa g_k}={\pa\over \pa g_k}{\pa\over \pa x}$,
${\pa\over \pa g_j}{\pa\over \pa g_k}={\pa\over \pa g_k}{\pa\over \pa g_j}$, as well as compatibility with the discrete recursion, imply {\bf integrability},
and allows to define a $\tau$-function \cite{MiwaJimbo, BEHtauiso}.

\bigskip
We define the spectral curve as the locus of eigenvalues of $\calD_n(x)$:
\beq
E_n(x,y):=\det(y\1 - \calD_n(x))
\eeq

\br
In the 1-hermitean-matrix model, $\calD_n$ is a $2\times 2$ matrix, and thus $\deg_y E_n(x,y)=2$, i.e. the curve $E_n(x,y)=0$ is an {\bf hyperelliptical curve}.
In other matrix models, one gets algebraic curves which are not hyperelliptical.
\er

\br
What we will se below, is that the curve $E_N(x,y)$ has a large $N$ limit $E(x,y)$, which is also an hyperelliptical curve.
In general, the matrix $\calD_N(x)$ has no large $N$ limit.
\er

\section{Riemann-Hilbert problems and isomonodromies}

The $2\times2$ system $\calD_N$ has $2$ independent solutions:
\beq
-{1\over N} \,{\pa \over \pa x} \pmatrix{\psi_n(x) \cr \psi_{n-1}(x)} = {\calD}_n(x)\,\pmatrix{\psi_n(x) \cr \psi_{n-1}(x)}
\virg
-{1\over N} \,{\pa \over \pa x} \pmatrix{\phi_n(x) \cr \phi_{n-1}(x)} = {\calD}_n(x)\,\pmatrix{\phi_n(x) \cr \phi_{n-1}(x)}
\eeq
where the wronskian is non-vanishing:  $\det\pmatrix{\psi_n(x) & \phi_n(x) \cr \psi_{n-1}(x) &  \phi_{n-1}(x)}\neq 0$.

We define the matrix of fundamental solutions:
\beq
\Psi_n(x):=\pmatrix{\psi_n(x) & \phi_n(x) \cr \psi_{n-1}(x) &  \phi_{n-1}(x)}
\eeq
it obeys the same differential equation:
\beq
-{1\over N} \,{\pa \over \pa x} \Psi_n(x) = {\calD}_n(x)\,\Psi_n(x)
\eeq

\medskip
Here, the second solution can be constructed explicitely:
\beq
\phi_n(x) = \e^{+{N\over 2}V(x)}\,\int {dx'\over x-x'}\,\psi_n(x') \e^{-{N\over 2}V(x')}
\eeq
Notice that $\phi_n(x)$ is discontinuous along the integration path of $x'$ (i.e. the real axis in the most simple case), the discontinuity is simply $2i\pi \psi_n(x)$.
In terms of fundamental solutions, one has the jump relation:
\beq\label{JumpRH}
\Psi_n(x+i0) = \Psi_n(x-i0)\,\pmatrix{1 & 2i\pi \cr 0 & 1}
\eeq

Finding an invertible piecewise analytical matrix, with given large $x$ behaviours, with given jumps on the borders between analytical domains, is called a {\bf Riemann--Hilbert problem} \cite{BlIt, BlIt1, BEHRH}.

It is known that the Riemann--Hilbert problem has a unique solution, and that if two R-H problems differ by $\epsilon$ (i.e. the difference between jumps and behaviours at $\infty$ is bounded
by $\epsilon$), then the two solutions differ by at most $\epsilon$ (roughly speeking, harmonic functions have their extremum on the boundaries).
Thus, this approach can be used \cite{BlIt, dkmvz, dkmvz2} in order to find large $N$ asymptotics of orthogonal polynomials:
The authors of \cite{BlIt} considered a guess for the asymptotics, which satisfies another R-H problem, which differs from this one by $O(1/N)$.

\bigskip
Notice that the jump matrix in \ref{JumpRH} is independent of $x$, of $n$ and of the potential, it is a constant.
The jump matrix is also called a monodromy, and the fact that the monodromy is a constant, is called {\bf isomonodromy} property \cite{MiwaJimbo}.

Consider an invertible, piecewise analytical matrix $\Psi_n(x)$, with appropriate behaviours\footnote{The behaviours at $\infty$ are far beyond the scope of this short lecture.
They are easily obtained by computing $\phi_n(x)$ by saddle point method at large $x$.} at $\infty$,
which satisfies \ref{JumpRH}, then, it is clear that the matrix
$-{1\over N} \Psi_n'(x) (\Psi_n(x))^{-1}$, has no discontinuity, and given its behaviour at $\infty$, it must be a polynomial.
Thus, we can prove that $\Psi_n(x)$ must satisfy a differential system $\calD_n(x)$ with polynomial coefficients.
Similarly, the fact that the monodromy is independent of $g_k$ and $n$ implies the deformation equations, as well as the discrete recursion relations.

Thus, the isomonodromy property, implies the existence of compatible differential systems, and integrability \cite{BI, FIK, stringIts, MiwaJimbo, TW2, BEHtauiso}.

\section{WKB--like asymptotics and spectral curve}
\label{secasympWKBformal}

Let us look for a formal solution of the form:
\beq\label{asympWKBformal}
\Psi_N(x) =  A_N(x) \, \e^{-N T(x)} B_N
\eeq
where $T(x)={\rm diag}(T_1(x),T_2(x))$ is a diagonal matrix, and $B_N$ is independent of $x$.
The differential system $\calD_N(x)$ is such that:
\bea
\calD_N(x) &=& -{1\over N}\Psi_N'\Psi_N^{-1} = A_N(x) T'(x) A_N^{-1}(x) - {1\over N} A'_N(x) A_N^{-1}(x) \cr
& =&  A_N(x) T'(x) A_N^{-1}(x) + O({1\over N})
\eea
this means, that, under the asumption that $A_N(x)$ has a large $N$ limit $A(x)$, $T'_1(x)$ and $T'_2(x)$ are the large $N$ limits of the eigenvalues of $\calD_N(x)$.

With such an hypothesis, one gets for the orthogonal polynomials:
\beq
\psi_N(x) \sim A_{11}\e^{-NT_1(x)} B_{1,1} + A_{12}\e^{-NT_2(x)} B_{2,1}
\eeq

We are now going to show how to derive such a formula.

\section{Orthogonal polynomials as matrix integrals}

\subsection{Heine's formula}

\bt Heine's theorem \cite{Szego}.
The orthogonal polynomials $p-n(x)$ are given by:
\bea
p_n(\xi) &=& {\int dx_1\dots dx_N\,\, \prod_{i=1}^N (\xi-x_i)\,\, (\Delta(x_1,\dots,x_N))^2\,\, \prod_{i=1}^N \e^{-NV(x_i)}\over
\int dx_1\dots dx_N\,\,  (\Delta(x_1,\dots,x_N))^2\,\, \prod_{i=1}^N \e^{-NV(x_i)}} \cr
&=& \left<\det(\xi\1 - M)\right>
\eea
\et
i.e. the orthogonal polynomial is the average of the characteristic polynomial of the random matrix.

Thus, we can define the orthogonal polynomials as matrix integrals, similar to the partition function $Z$ define in \ref{defZ}.

\subsection{Another matrix model}

Define the potential:
\beq
V_h(x):=V(x)-h\ln{(\xi-x)}
\eeq
and the partition function:
\beq
Z_n(h,T):=\e^{-{n^2\over T^2}F_n(h,T)}
:=\int dx_1\dots dx_n\,\,  (\Delta(x_1,\dots,x_n))^2\,\, \prod_{i=1}^n \e^{-{n\over T}V_h(x_i)}
\eeq
i.e. $Z_N(0,1)=Z$ is our initial partition function.

Heine's formula reads:
\beq
p_n(\xi) = {Z_n({1\over N},{n\over N})\over Z_n(0,{n\over N})}
= \e^{-N^2(F_n({1\over N},{n\over N})-F_n(0,{n\over N}))}
\eeq

The idea, is to perform a Taylor expansion in $h$ close to $0$ and $T$ close to $1$.

\subsubsection{Taylor expansion}

We are interested in $n=N$ and $n=N-1$, thus $T={n\over N}=1+{n-N\over N}=1+O(1/N)$ and $h=0$ or $h=1/N$, i.e. $h=O(1/N)$:
\beq
T=1+O(1/N)
\virg
h=O(1/N)
\eeq

Roughly speaking:
\bea\label{asymp1}
p_n(\xi)
&\sim& \e^{-N^2\left( h {\pa F\over \pa h}+(T-1)h{\pa^2 F\over \pa h\pa T}+{h^2\over 2}{\pa^2 F\over \pa h^2}+O(1/N^3)\right)} \cr
&\sim& \e^{-N {\pa F\over \pa h}}\,\e^{-(n-N){\pa^2 F\over \pa h\pa T}}\,\e^{-{1\over 2}{\pa^2 F\over \pa h^2}}\,\,(1+O(1/N))
\eea
where all the derivatives are computed at $T=1$ and $h=0$.

\subsubsection{Topological expansion}

Imagine that $F_n$ has a $1/n^2$ expansion of the form:
\beq
F = F^{(0)} + {1\over n^2} F^{(1)} + O({1\over n^3})
\eeq
where all $F^{(0)}$ and $F^{(1)}$ are analytical functions of $T$ and $h$, than one needs only $F^{(0)}$ in order to compute the asymptotics \ref{asymp1}.

\smallskip

Actualy, that hypothesis is not always true. It is wrong in the so called ''mutlicut'' case.
But it can be adapted in that case, we will come back to it in section \ref{sectmulticutasymp}.
For the moment, let us conduct the calculation only with $F^{(0)}$.

\section{Computation of derivatives of $F^{(0)}$}

We have defined:
\beq
Z_n(h,T)=\e^{-{n^2\over T^2}F_n(h,T)}
= \int dM_{n\times n} \e^{-{n\over T}\tr V(M)}\, \e^{h{n\over T}\ln{(\xi-M)}}
\eeq
this implies that:
\beq
-{n^2\over T^2} {\pa F_n\over \pa h} = \left<{n\over T}\tr \ln{(\xi-M)}\right>_{V_h}
\eeq
i.e.
\bea
{\pa F_n\over \pa h}
&=& -{T\over n}\left<\tr \ln{(\xi-M)}\right>_{V_h} \cr
\eea
It is  a primitive of $-{T\over n}\left<\tr \ln{(x-M)}\right>_{V_h}$, which behaves as $-{T\over n}\ln{x}+O(1/x)$ at large $x$.
Therefore, we define the resolvent $W(x)$:
\beq\label{defWVh}
W(x):={T\over n}\left<\tr {1\over x-M}\right>_{V_h}
\eeq
Notice that it depends on $\xi$ through the potential $V_h$, i.e. through the average $<.>$.
And we define the effective potential:
\beq
{V_{\rm eff}}(x)= V_h(x)-2T\ln{x}-2\int_{\infty}^x (W(x')-{T\over x'}) dx'
\eeq
which is  a primitive of $V'_h(x)-2W(x)$.
Thus , we have:
\beq
{\pa F_n\over \pa h} = {1\over 2}\left({V_{\rm eff}}(\xi)-V_h(\xi)\right)
\eeq

We also introduce:
\beq\label{defOm}
\Omega(x):={\pa W(x)\over \pa T}
\virg
\ln{\L(x)}:=\ln{x}+\int_{\infty}^x(\Omega(x')-{1\over x'})dx' = -{1\over 2}{\pa \over \pa T}V_{\rm eff}(x)
\eeq

\beq\label{defH}
H(x,\xi):={\pa W(x)\over \pa h}
\virg
\ln{H(\xi)}:=\int_{\infty}^\xi H(x',\xi)dx'
\eeq

i.e.
\beq
{\pa^2 F_n\over \pa h^2} = -\ln{H(\xi)}
\virg
{\pa^2 F_n\over \pa h\pa T} = -\ln{\L(\xi)}
\eeq

With these notations, the asymptotics are:
\beq
\psi_n(\xi)
\sim
\sqrt{H(\xi)}\,\,\left(\L(\xi)\right)^{n-N}\,\,\e^{-{N\over 2}{V_{\rm eff}}(\xi)}\,\,(1+O(1/N))
\eeq

Now, we are going to compute $W$, $\L$, $H$, etc, in terms of geometric properties of an hyperelliptical curve.

\br
This is so far only a sketch of the derivation, valid only in the 1-cut case.
In general, $F_n$ has no $1/n^2$ expansion, and that case will be addressed in section \ref{sectmulticutasymp}.
\er

\br
These asymptoics are of the form of \ref{asympWKBformal} in section.\ref{secasympWKBformal}, and thus,
${1\over 2}V'(x)-W(x)$ is the limit of the eigenvalues of $\calD_N(x)$.
\er

\section{Saddle point method}

There exists many ways of computing the resolvent and its derivatives with respect to $h$, $T$, or other parameters.
The loop equation method is a  very good method, but there is not enough time to present it here.
There are several saddle-point methods, which all coincide to leading order.
We are going to present one of them, very intuitive, but not very rigorous on a mathematical ground, and not very appropriate for next to leading computations.
However, it gives the correct answer to leading order.

\bigskip

Write:
\beq
Z_n(h,T)=\e^{-{n^2\over T^2}F_n(h,T)}=\int dx_1\dots dx_n \e^{-{n^2\over T^2}{\cal S}(x_1,\dots,x_n)}
\eeq
where the action is:
\beq
{\cal S}(x_1,\dots,x_n):={T\over n}\sum_{i=1}^n V_h(x_i) -2{T^2\over n^2}\sum_{i>j} \ln{(x_i-x_j)}
\eeq

The saddle point method consists in finding configurations $x_i=\xbar_i$ where ${\cal S}$ is extremal,
i.e.
\beq
\forall i=1,\dots n, \qquad \left.{\pa {\cal S}\over \pa x_i}\right|_{x_j=\xbar_j}=0
\eeq
i.e., we have the {\bf saddle point equation}:
\beq
\forall i=1,\dots n, \qquad
V'_h(\xbar_i) = 2{T\over n}\sum_{j\neq i} {1\over \xbar_i-\xbar_j}
\eeq

The saddle point approximation\footnote{The validity of the saddle point approximation is not proven rigorously  for large number of variables.
But here, we have many evidences that we can trust the results it gives. The asymptotics we are going to find have been proven rigorously by other methods.
Basicaly, it is expected to work because the number of variables $n$ is small compared to the large parameter $n^2$ in the action.}
consists in writting:
\beq
Z_n(h,T) \sim {1\over \sqrt{\det\left(\pa {\cal S}\over \pa x_i\pa x_j\right)}}\,\,\e^{-{n^2\over T^2}{\cal S}(\xbar_1,\dots,\xbar_n)}\,\,(1+O(1/n))
\eeq
where $(\xbar_1,\dots,\xbar_n)$ is the solution of the saddlepoint equation which minimizes $\Re {\cal S}$.

\br
The saddle point equation may have more than one minimal solution $(\xbar)$.

- in particular if $\xi\in \R$, there are two solutions, complex conjugate of each other.

- in the multicut case, there are many saddlepoints with near-minimal action.

In all cases, one needs to sum over all the saddle points.
Let us call $\{\xbar\}_I$, the collection of saddle points. We have:
\beq
Z_n \sim \sum_I {C_I\over \sqrt{{\cal S}''(\{\xbar\}_I)}}\,\,\e^{-{n^2\over T^2}{\cal S}(\{\xbar\}_I)}\,\,(1+O(1/n))
\eeq
Each saddle point $\{\xbar\}_I$ corresponds  to a particular minimal $n$-dimensional integration path in $\C^n$,noted $\Gamma_I$,
and the coefficients $C_I\in \Z$ are such that:
\beq
\R^n = \sum_I C_I \Gamma_I
\eeq
\er

\section{Solution of the saddlepoint equation}

We recall the saddle point equation:
\beq\label{sadlepointxbar}
\forall i=1,\dots n, \qquad
V'_h(\xbar_i) = 2{T\over n}\sum_{j\neq i} {1\over \xbar_i-\xbar_j}
\eeq
We introduce the function:
\beq\label{defom}
\om(x):={T\over n}\sum_{j=1}^n {1\over x-\xbar_j}
\eeq
in the large $N$ limit, $\om(x)$ is expected to tend toward the resolvent, at least in the case there is only one minimal saddle point.
Indeed, the $\xbar_i$'s are the position of the eigenvalues minimizing the action, i.e. the most probable positions of eigenvalues of $M$, and thus
\ref{defom} should be close to ${T\over n}\tr {1\over x-M}$.

\subsection{Algebraic method}

Compute $\om^2(x)+{T\over n}\om'(x)$, you find:
\bea
\om^2(x)+{T\over n}\om'(x)
&=& {T^2\over n^2} \sum_{i=1}^n \sum_{j=1}^n {1\over (x-\xbar_i)(x-\xbar_j)} - {T^2\over n^2} \sum_{i=1}^n {1\over (x-\xbar_i)^2} \cr
&=&{T^2\over n^2} \sum_{i\neq j}^n {1\over (x-\xbar_i)(x-\xbar_j)} \cr
&=&{T^2\over n^2} \sum_{i\neq j}^n \left({1\over x-\xbar_i}-{1\over x-\xbar_j}\right)\,{1\over \xbar_i-\xbar_j} \cr
&=&{2T^2\over n^2} \sum_{i=1}^n {1\over x-\xbar_i}\,\sum_{j\neq i}^n {1\over \xbar_i-\xbar_j} \cr
&=&{T\over n} \sum_{i=1}^n {V'_h(\xbar_i)\over x-\xbar_i} \cr
&=&{T\over n} \sum_{i=1}^n {V'_h(x)-(V'_h(x)-V'_h(\xbar_i))\over x-\xbar_i} \cr
&=& V'_h(x)\om(x)-{T\over n} \sum_{i=1}^n {V'_h(x)-V'_h(\xbar_i)\over x-\xbar_i} \cr
&=&(V'(x)-{h\over x-\xi})\om(x)-{T\over n} \sum_{i=1}^n {V'(x)-V'(\xbar_i)\over x-\xbar_i} + h{\om(\xi)\over x-\xi}\cr
\eea
i.e. we get the equation:
\beq
\om^2(x)+{T\over n}\om'(x)
= V'(x)\om(x)- P(x) -  h{\om(x)-\om(\xi)\over x-\xi}
\eeq
where $P(x):={T\over n} \sum_{i=1}^n {V'(x)-V'(\xbar_i)\over x-\xbar_i}$ is a polynomial in $x$ of degree at most $\deg V-2$.

In the large $N$ limit, if we assume\footnote{It is possible to do the calculation without droping the $1/N$ term. One gets a Ricati equation, which is equivalent to a Schroedinger equation.
If one is interested in a large N limit for the resolvent, the asymptotic analysis of that Schroedinger equation (Stokes phenomenon) gives, to leading order, the same thing as when one drops the $1/N$ term. If one whishes to go beyond leading order, many subtleties occur.} that we can drop the $1/N W'(x)$ term, we get an algebraic equation, which is in this case
an hyperelliptical curve.
In particular at $h=0$ and $T=1$:
\beq
\om(x) = {1\over 2}\left(V'(x)-\sqrt{V'^2(x)-4P(x)}\right)
\eeq

The properties of this algebraic equation have been studied by many authors, and the $T$ and $h$ derivatives, as well as other derivatives were computed in various works.
Here, we briefly sketch the method.
See \cite{kriechever, KazMar, eynmultimat} for more details.

\subsection{Linear saddle point equation}

In the large $N$ limit, both the average density of eigenvalues, and the density of $\xbar$ tend towards a continuous compact support density $\rhobar(x)$.
In that limit, the resolvent is given by:
\beq
\om(x) = T \, \int_{{\rm supp}\,\,\rhobar}{\rhobar(x')\,dx'\over x-x'}
\eeq
i.e.
\beq
\forall x\in{\rm supp}\,\,\rhobar, \qquad
\rhobar(x) = -{1\over 2i\pi T}(\om(x+i0)-\om(x-i0))
\eeq
and the saddle point equation \ref{sadlepointxbar}, becomes a linear functional equation:
\beq\label{sadlepointequrho}
\forall x\in{\rm supp}\,\,\rhobar, \qquad
V'_h(x) = \om(x+i0)+\om(x-i0)
\eeq

The advantage of that equation, is that it is linear in $\om$, and thus in $\rhobar$.
The nonlinearity is hidden in ${\rm supp}\,\,\rhobar$.

\subsubsection{Example: One cut}

If the support of $\rhobar$ is a single interval:
\beq
{\rm supp}\,\,\rhobar = [a,b]\virg a<b
\eeq
then, look for a solution of the form:
\beq
\om(x) = {1\over 2}\left(V'_h(x) - M_h(x)\sqrt{(x-a)(x-b)}\right)
\eeq
The saddle point equation \ref{sadlepointequrho} implies that $M_h(x+i0)=M_h(x-i0)$, i.e. $M_h$ has no discontinuities,
and because of its large $x$ behaviour, as well as its behaviours near $\xi$, it must be a rational function of $x$, with a simple pole at $x=\xi$.
$M_h$, $a$ and $b$ are entirely determined by their behaviours near poles, i.e.:
\beq
\om(x) \mathop\sim_{x\to\infty} {T\over x}
\eeq
\beq
\om(x) \mathop\sim_{x\to\xi} {\rm regular}\quad \longrightarrow M_h(x)\mathop\sim_{x\to\xi} -{h\over x-\xi}
\eeq
Thus, one may write:
\beq
\om(x) = {1\over 2}\left(V'(x) - M(x)\sqrt{(x-a)(x-b)} - {h\over x-\xi}\left(1-{\sqrt{(x-a)(x-b)}\over \sqrt{(\xi-a)(\xi-b)}}\right)\right)
\eeq
where $M(x)$ is now a polynomial (which still depends on $h$ and $T$ and the other parameters), it is such that:
\beq
M(x) = \mathop{\rm Pol}_{x\to\infty}\,\, {V'(x)\over \sqrt{(x-a)(x-b)}}
\eeq
The density is thus:
\beq
\rhobar(x) ={1\over 2\pi T}M_h(x)\sqrt{(x-a)(b-x)}
\virg
{\rm supp}\,\,\rhobar = [a,b]
\eeq

$$\begin{array}{r}
{\epsfxsize 12cm\epsffile{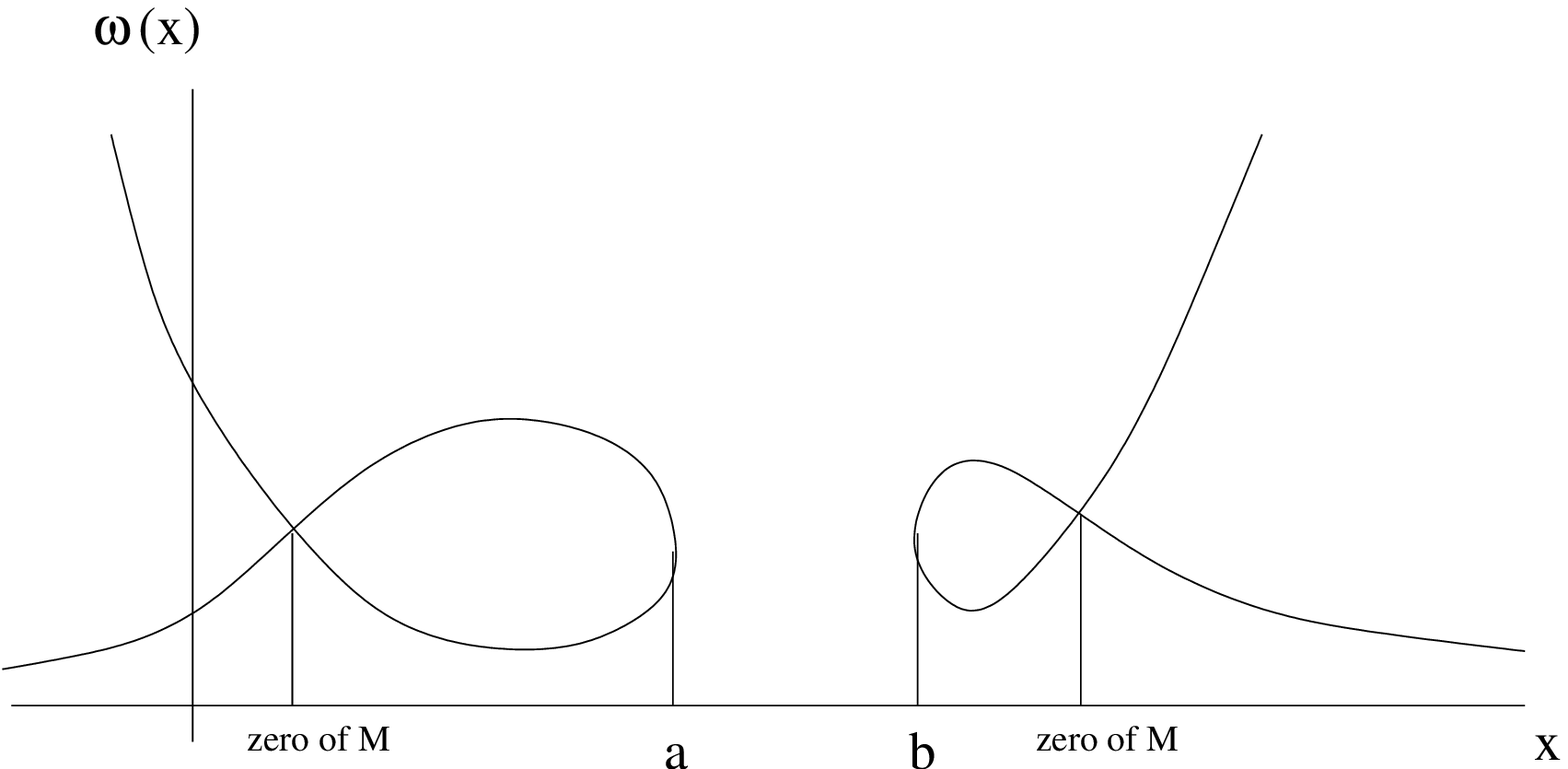}}
\end{array}$$

\subsubsection{Multi-cut solution}

Let us assume that the support of $\rhobar$ is made of $s$ separated intervals:
\beq
{\rm supp}\,\,\rhobar = \cup_{i=1}^s [a_i,b_i]
\eeq
then, for any sequence of integers $n_1,n_2,\dots, n_s$ such that $\sum_{i_1}^s n_i=n$, it is possible
to find a solution for the saddle point equation.
That solution obeys \ref{sadlepointequrho}, as well as the conditions:
\beq
\forall i=1,\dots,s
\virg
\int_{a_i}^{b_i} \rho(x) dx = T {n_i\over N}
\eeq

The solution of the saddle point equation can be described as follows:

let the polynomial $\sigma(x)$ be defined as:
\beq
\sigma(x):=\prod_{i=1}^s (x-a_i)(x-b_i)
\eeq
The solution of the saddle point equation \ref{sadlepointequrho}, is of the form:
\beq
\om(x) = {1\over 2}\left(V'_h(x) - M_h(x)\sqrt{\sigma(x)}\right)
\eeq
where $M_h(x)$ is a rational function of $x$, with a simple pole at $x=\xi$.
$M_h$, and $\sigma(x)$ are entirely determined by their behaviours near poles, i.e.:
\beq
\om(x) \mathop\sim_{x\to\infty} {T\over x}
\eeq
\beq
\om(x) \mathop\sim_{x\to\xi} {\rm regular}\quad \longrightarrow M_h(x)\mathop\sim_{x\to\xi} -{h\over x-\xi}
\eeq
and by the conditions that:
\beq
\forall i=1,\dots,s \virg \int_{a_i}^{b_{i}} M_h(x)\sqrt{\sigma(x)} dx = 2i\pi T{n_i\over n}
\eeq

\subsection{Algebraic geometry: hyperelliptical curves}

Consider the curve given by:
\beq
\om(x) = {1\over 2}\left(V_h'(x) - M_h(x)\sqrt{(x-a)(x-b)}\right)
\eeq
It has two sheets, i.e. for each $x$, there are two values of $\om(x)$, depending on the choice of sign of the square-root.

- In the physical sheet (choice $+\sqrt{}$), it behaves near $\infty$ like $\om(x)\sim T/x$

- In the second sheet (choice $-\sqrt{}$), it behaves near $\infty$ like $\om(x)\sim V'_h(x)$

Since $\om(x)$ is a complex valued, analytical function of a cmplex variable $x$,
the curve can be thought of as the embedding of a Riemann surface into $\C\times \C$.

I.e. we have a Riemann surface $\curve$, with two (monovalued) functions defined on it:
$p\in\curve\, , \,\, \to x(p)\in\C$, and $p\in\curve\, , \,\, \to \om(p)\in\C$.
For each $x$, there are two $p\in\curve$ such that $x(p)=x$, and this is why there are two values of $\om(x)$.

Each of the two sheets is homeomorphic to the complex plane, cut along the segments $[a_i,b_i]$, and the two sheets are glued together along the cuts.
The complex plane, plus its point at infinity, is the Riemann sphere.
Thus, our curve $\curve$, is obtained by taking two Riemann spheres, glued together along $s$ circles.
It is a genus $s-1$ surface.

$$\begin{array}{r}
{\epsfxsize 14cm\epsffile{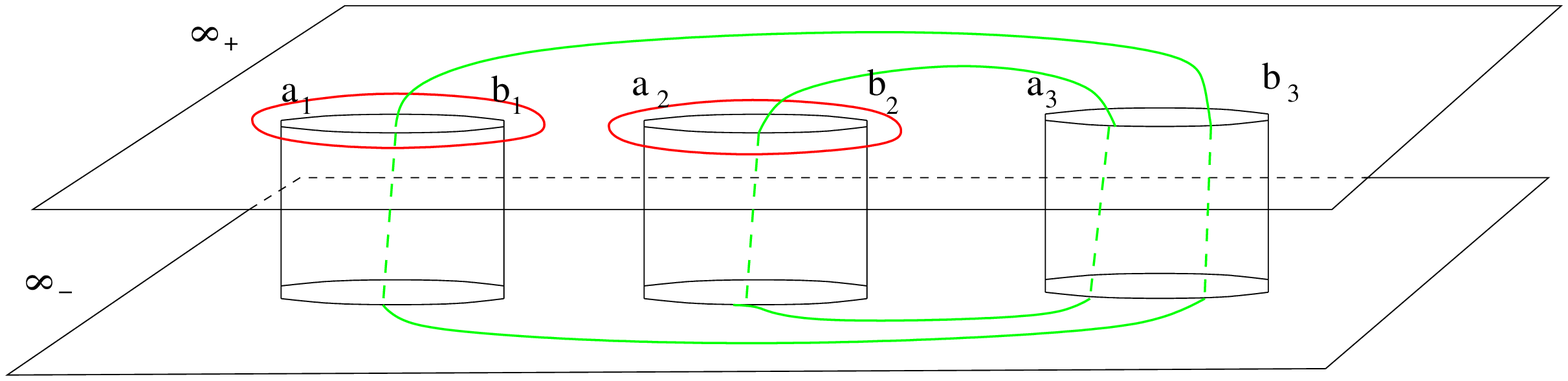}}
\end{array}$$

$$\begin{array}{r}
{\epsfysize 4cm\epsffile{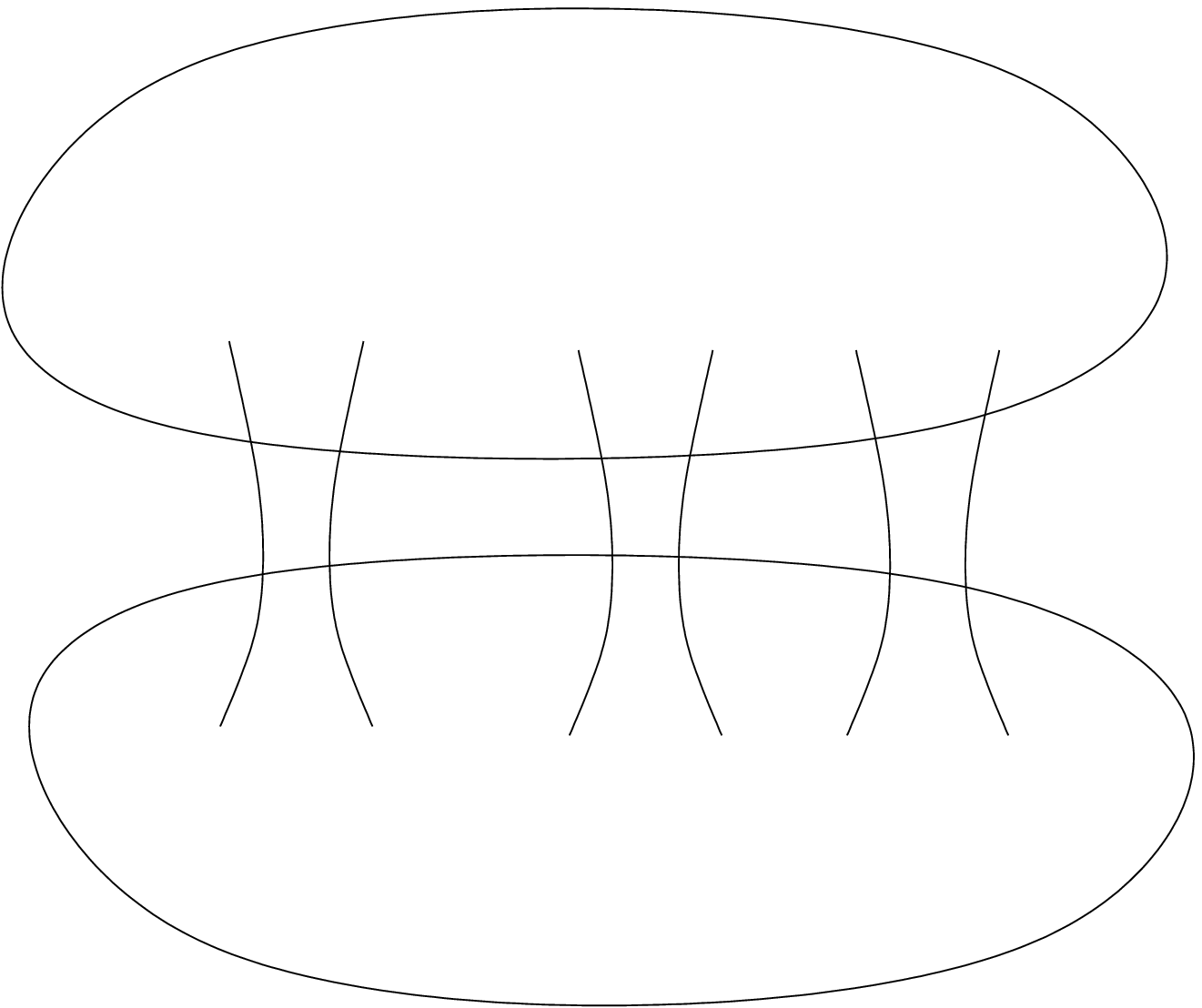}}
\end{array}$$

\subsection{Genus zero case (one cut)}

If the curve as genus zero, it is homeomorphic to the Riemann sphere $\curve=\C$.
One can always choose a rational parametrization:
\beq
x(p)={a+b\over 2}+\gamma(p+{1/p})
\virg \gamma={b-a\over 4}
\eeq
\beq
\sqrt{(x-a)(x-b)}=\gamma(p-1/p)
\eeq
so that $\om$ is a rational function of $p$.

That representation maps the physical sheet onto the exterior of the unit circle, and the second sheet onto the interior of the unit circle.
The unit circle is the image of the two sides of the cut $[a,b]$, and the branchpoints $[a,b]$ are maped to $-1$ and $+1$.
Changing the sign of the square root is equivalent to changing $p\to 1/p$.

The branch points are of course the solutions of $dx/dp=0$, i.e. $dx(p)=0$:
\beq
dx(p) = \gamma\,\left(1-{1\over p^2}\right)\, dp
\virg
dx(p)=0\leftrightarrow p=\pm 1 \leftrightarrow x(p)=a,b
\eeq

There are two points at $\infty$, $p=\infty$ in the physical sheet, and $p=0$ in the second sheet.

$$\begin{array}{r}
{\epsfxsize 9cm\epsffile{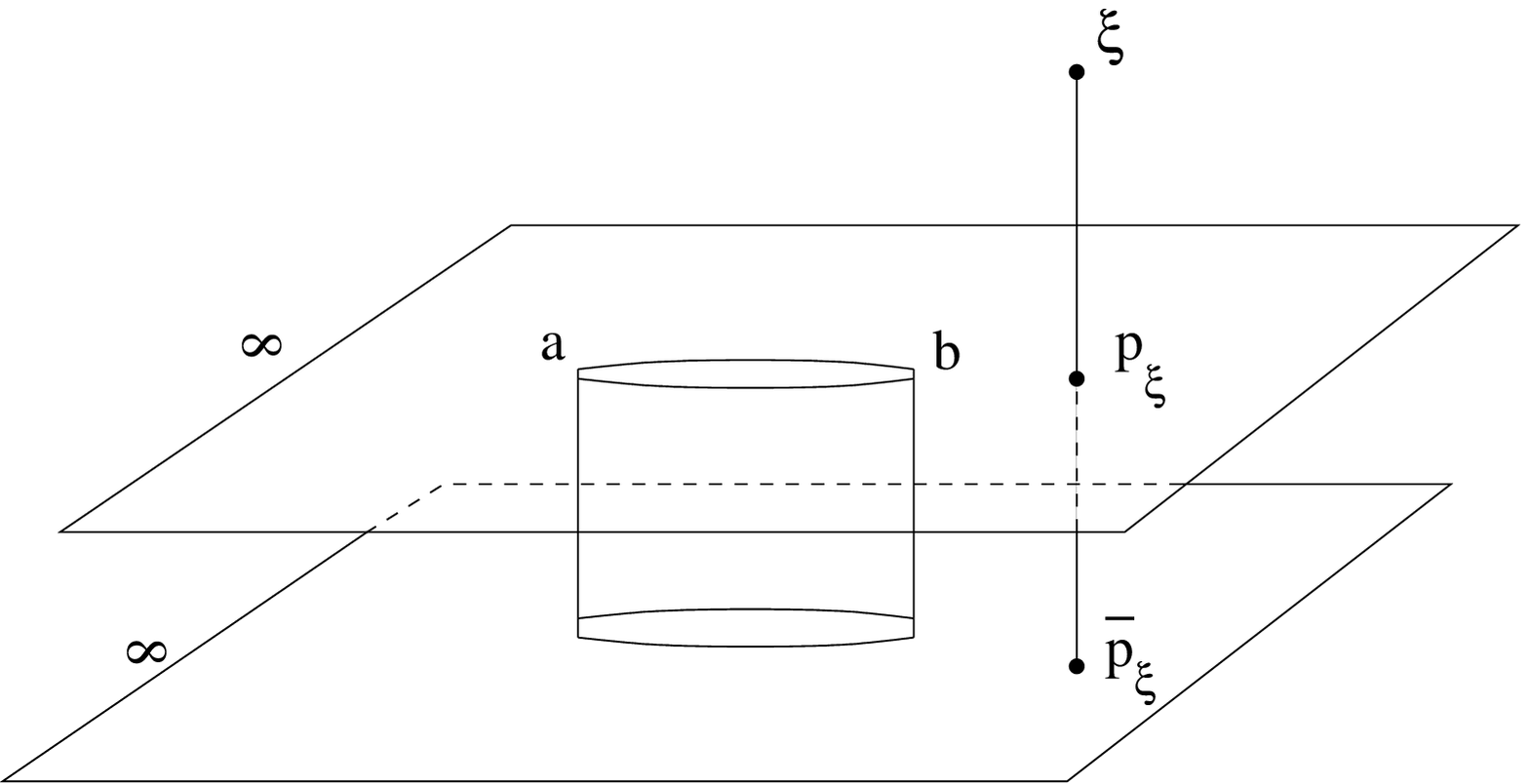}}
\,\,
{\epsfxsize 7cm\epsffile{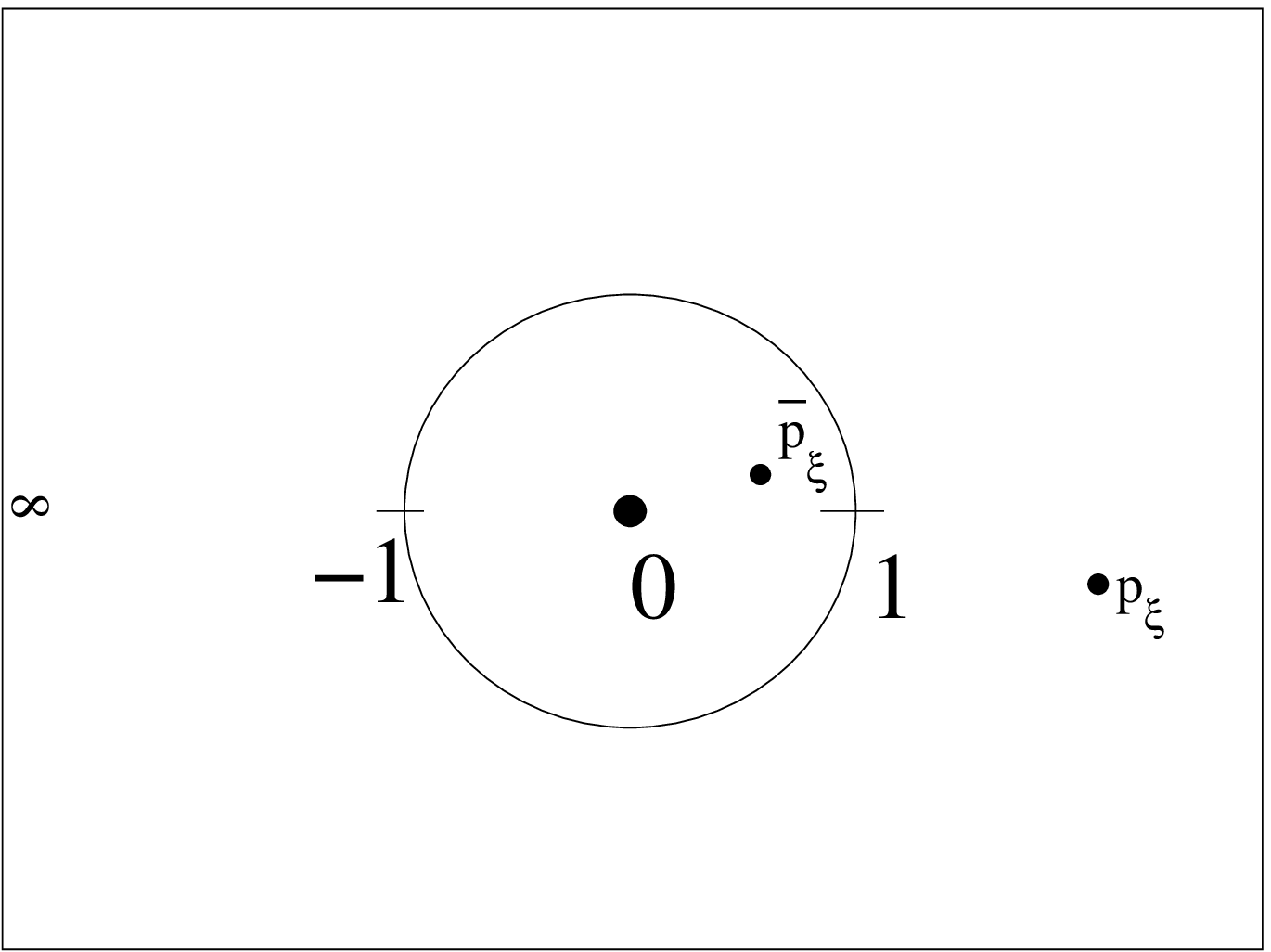}}
\end{array}$$

\bigskip

Since the resolvent $\om(p)$ is a rational function of $p$, it is then entirely determined by its behaviour near its poles.
the poles are at $p=\infty$, $p=0$,
$p=p_\xi$ and $p=\pbar_\xi$ (the two points of $\curve$ such that $x(p)=\xi$, such that $p_\xi$ is in the physical sheet, and $\pbar_\xi$ is in the second sheet):
The boundary conditions:
\beq\label{bndomgzero}
\left\{
\begin{array}{l}
\displaystyle \om(p) \mathop\sim_{p\to\infty} {T\over x(p)} \cr
\displaystyle \om(p) \mathop\sim_{p\to 0} V'(x(p))-{T\over x(p)}-{h\over x(p)} \cr
\displaystyle \om(p) \mathop\sim_{p\to \pbar_\xi} -{h\over x(p)-\xi} \cr
\displaystyle \om(p) \mathop\sim_{p\to p_\xi} {\rm regular} \cr
\end{array}
\right.
\eeq

\subsubsection{$T$ derivative}

Now, let us compute  $\pa\om(p)/\pa T$ at $x(p)$ fixed.
Eq. \ref{bndomgzero} becomes:
\beq
\left\{
\begin{array}{l}
\displaystyle {\pa \om(p)\over \pa T} \mathop\sim_{p\to\infty} {1\over x(p)} \cr
\displaystyle {\pa \om(p)\over \pa T} \mathop\sim_{p\to 0} -{1\over x(p)} \cr
\displaystyle {\pa \om(p)\over \pa T} \mathop\sim_{p\to \pbar_\xi} {\rm regular} \cr
\displaystyle {\pa \om(p)\over \pa T} \mathop\sim_{p\to p_\xi} {\rm regular} \cr
\end{array}
\right.
\eeq
Moreover, we know that $\om(x)$ has a square-root behaviour near $a$ and $b$, in $\sqrt{(x-a)(x-b)}$, and $a$ and $b$ depend on $T$,
thus $\pa\om/\pa T$ may behave in $((x-a)(x-b))^{-1/2}$ near $a$ and $b$, i.e. $\pa\om/\pa T$ may have simple poles at $p=\pm 1$.

Finaly, $\pa\om(p)/\pa T$, has simple poles at $p=1$ and $p=-1$, and vanishes at $p=0$ and $p=\infty$, the only possibility is:
\beq
\left.{\pa \om(p)\over \pa T}\right|_{x(p)}={p\over \gamma (p^{2}-1)} = {1\over p}\, {dp\over dx}
\eeq
which is better written in terms of differential forms:
\beq
\left.{\pa \om(p)\over \pa T}\right|_{x(p)}\, dx(p)= {dp\over p}=d\ln{p}
\eeq
the RHS is independent of the potential, it is universal.

With the notation \ref{defOm}, we  have:
\beq
\Om(p)dx(p)={dp\over p}
\virg
\L(p)=\gamma p
\eeq

\subsubsection{$h$ derivative}

The $h$ derivative is computed in a very similar way.
\beq
\left\{
\begin{array}{l}
\displaystyle {\pa \om(p)\over \pa h} \mathop\sim_{p\to\infty} O(p^{-2}) \cr
\displaystyle {\pa \om(p)\over \pa h} \mathop\sim_{p\to 0} -{1\over x(p)} \cr
\displaystyle {\pa \om(p)\over \pa h} \mathop\sim_{p\to \pbar_\xi} -{1\over x(p)-\xi} \cr
\displaystyle {\pa \om(p)\over \pa h} \mathop\sim_{p\to p_\xi} {\rm regular} \cr
\end{array}
\right.
\eeq
implies that $\pa \om/\pa h$ can have poles at $p=\pm 1$ and at $p=\pbar_\xi$, and vanishes at $p=0$.
The only possibility is:
\beq
\left.{\pa \om(p)\over \pa h}\right|_{x(p)}={-p\,\pbar_\xi\over \gamma (p-\pbar_\xi)(p^{2}-1)}
\eeq
i.e.
\beq
\left.{\pa \om(p)\over \pa h}\right|_{x(p)}\, dx(p) = {dp\over p}-{dp\over p-\pbar_\xi} = d\ln{p\over p-\pbar_\xi}
\eeq
which again is universal.

With the notation \ref{defH}, we  have:
\beq
H(p,p_\xi)dx(p)={dp\over p}-{dp\over p-{1\over p_\xi}}
\virg
H(p_\xi) = \ln{\left({p_\xi\over p_\xi-\pbar_\xi}\right)} = -\ln{\left({1\over\gamma}\,{dx\over dp}(\xi)\right)}
\eeq

\subsection{Higher genus}

For general genus, the curve can be parametrized by $\theta$-functions.
Like rational functions  for genus 0, $\theta$-functions are the building blocks of functions defined on a compact Riemann surface, and any such function is
entirely determined by its behaviour near its poles, as well as by its integrals around irreducible cycles.
All the previous paragraph can be extended to that case.

Let $\infty_+$ and $\infty_-$ be the points at infinity, i.e. the two poles of $x(p)$, with $\infty_+$ in the physical sheet and $\infty_-$ in the second sheet.
Let $p=p_\xi$ and $p=\pbar_\xi$ be the two points of $\curve$ such that $x(p)=\xi$, and with $p_\xi$ in the physical sheet, and $\pbar_\xi$ in the second sheet.

The differential form $\om(p) dx(p)$ is entirely determined by:
\beq\label{ompoles}
\left\{
\begin{array}{ll}
\displaystyle \om(p)dx(p) \mathop\sim_{p\to \infty_+} T\,{dx(p)\over x(p)}  & \displaystyle  \virg \mathop\Res_{\infty_+} \om(p)dx(p)=-T \cr
\displaystyle \om(p)dx(p) \mathop\sim_{p\to \infty_-} dV(x(p)) - T{dx(p)\over x(p)}-h{dx(p)\over x(p)} & \displaystyle \virg \mathop\Res_{\infty_-} \om(p)dx(p)=T+h \cr
\displaystyle \om(p)dx(p) \mathop\sim_{p\to \pbar_\xi} -h{dx(p)\over x(p)-\xi} & \displaystyle \virg \mathop\Res_{\pbar_\xi} \om(p)dx(p)=-h \cr
\displaystyle \om(p)dx(p) \mathop\sim_{p\to p_\xi} {\rm regular} & \displaystyle \virg \mathop\Res_{p_\xi} \om(p)dx(p)=0 \cr
\displaystyle \oint_{{\cal A}_i} \om(p) dx(p) = T{n_i\over n} = {n_i\over N}  & \cr
\end{array}
\right.
\eeq
Since $\pa\om /\pa T,h $ can diverge at most like $(x-a_i)^{-1/2}$ near a branch point $a_i$, and $dx(p)$ has a zero at $a_i$, the differential form
$\pa \om dx/\pa T,h$ has no pole at the branch points.

\subsection{Introduction to algebraic geometry}

We introduce some basic concepts of algebraic geometry. We refer the reader to \cite{Farkas, Fay} for instance.

\bt
Given two points $q_1$ and $q_2$ on the Riemann surface $\curve$, there exists a unique differential form $dS_{q_1,q_2}(p)$,
with only two simple poles, one at $p=q_1$ with residue $+1$ and one at $p=q_2$ with residue $-1$, and which is normalized on the ${\cal A}_i$ cycles, i.e.
\beq\label{defdS}
\left\{
\begin{array}{l}
\displaystyle  \mathop\Res_{p\to q_1} dS_{q_1,q_2}(p)=+1 \cr
\displaystyle  \mathop\Res_{p\to q_2} dS_{q_1,q_2}(p)=-1 \cr
\displaystyle \oint_{{\cal A}_i} dS_{q_1,q_2}(p) = 0 \cr
\end{array}
\right.
\eeq
$dS$ is called an ``abelian differential of the third kind''.
\et

Starting from the behaviours near poles and irreducible cycles \ref{ompoles}, we easily find:
\beq
\Om(p)dx(p)=\left.{\pa\om(p) dx(p)\over \pa T}\right|_{x(p)} = -dS_{\infty_+,\infty_-}(p)
\eeq
\beq
H(p,p_\xi) dx(p) = \left.{\pa\om(p) dx(p)\over \pa h}\right|_{x(p)} = -dS_{\pbar_\xi,\infty_-}(p) = dS_{p_\xi,\infty_+}(p)-d\ln{\left(x(p)-x(p_\xi)\right)}
\eeq

\bt
On an algebraic curve of genus $g$, there exist exactly $g$ linearly independent ``holomorphic differential forms'' (i.e. with no poles), $du_i(p)$, $i=1,\dots, g$.
They can be chosen normalized as:
\beq
\oint_{{\cal A}_i} du_j(p)=\delta_{ij}
\eeq
\et
For hyperelliptical surfaces, it is easy to see that if $L(x)$ is a polynomial of degree at most $g-1=s-2$, the differential form
${L(x)\over \sqrt{\prod_{i=1}^s (x-a_i)(x-b_i)}}dx$ is regular at $\infty$, at the branch points, and thus has no poles.
And there are $g$ linearly independent polynomials of degree at most $g-1$. The irreducible cycles ${\cal A}_i$ is a contour surrounding $[a_i,b_i]$ in the positive direction.

\bd The matrix  of periods is defined by:
\beq
\tau_{ij}:=\oint_{{\cal B}_i} du_j(p)
\eeq
where the irreducible cycles ${\cal B}_i$ are chosen canonicaly conjugated to the ${\cal A}_i$, i.e. ${\cal A}_i\cap{\cal B}_j=\delta_{ij}$.
In our hyperelliptical case, we choose ${\cal B}_i$ as a contour crossing $[a_i,b_i]$ and $[a_s,b_s]$.

The matrix of periods is symmetric $\tau_{ij}=\tau_{ji}$, and its imaginary part is positive $\Im\tau_{ij}>0$.
It encodes the complex structure of the curve.
\ed

The holomrphic forms naturaly define an embedding of the curve into $\C^g$:
\bd
Given a base point $q_0\in\curve$, we define the Abel map:
\bea
\curve &\longrightarrow& \C^g \cr
p &\longrightarrow& {\vec u}(p) = (u_1(p),\dots,u_g(p)) \virg u_i(p):=\int_{q_0}^p du_i(p)
\eea
where the integration path is chosen so that it does not cross any ${\cal A}_i$ or ${\cal B}_i$.
\ed

\bd
Given a symmetric matrix $\tau$ of dimension $g$, such that $\Im\tau_{ij}>0$, we define the $\theta$-function, from $\C^g\to \C$ by:
\beq\label{deftheta}
\theta(\vec{u},\tau) = \sum_{\vec{m}\in \Z^g} \e^{i\pi {\vec m}^t \tau\vec{m}}\,\e^{2i\pi {\vec m}^t\vec{u}}
\eeq
It is an even entire function.
For any $\vec{m}\in \Z^g$, it satisfies:
\beq
\theta(\vec{u}+\vec{m})=\theta(\vec{u})
\virg
\theta(\vec{u}+\tau\vec{m})=\e^{-i\pi(2 {\vec m}^t\vec{u} + {\vec m}^t\tau \vec{m})}\, \theta(\vec{u})
\eeq
\ed

\bd The theta function vanishes on a codimension $1$ submanifold of $\C^g$, in particular, it vanishes at the odd half periods:
\beq
\vec{z}={{\vec m}_1+\tau\, {\vec m}_2\over 2}
\,\, ,\,\,\,
{\vec m}_1\in \Z^g\, ,\,\, {\vec m}_2\in \Z^g \,\, , \,\,\, ({\vec m}_1^t{\vec m}_1)\in 2\Z+1
\,\,\,\longrightarrow\,\,
\theta(\vec{z})=0
\eeq
For a given such odd half-period, we define the characteristic $\vec{z}$ $\theta$-function:
\beq
\theta_{\vec{z}}(\vec{u}):=\e^{i\pi m_2\vec{u}+}\,\theta(\vec{u}+\vec{z})
\eeq
so that:
\beq
\theta_{\vec{z}}(\vec{u}+\vec{m})= \e^{i\pi \vec{m}_2^t\vec{m}}\,\theta_{\vec{z}}(\vec{u})
\virg
\theta_{\vec{z}}(\vec{u}+\tau\vec{m})= \e^{-i\pi \vec{m}_1^t \vec{m}}\,\e^{-i\pi(2 {\vec m}^t\vec{u} + {\vec m}^t\tau \vec{m})}\, \theta_{\vec{z}}(\vec{u})
\eeq
and
\beq
\theta_{\vec{z}}(\vec{0})=0
\eeq
\ed

\bd
Given two points $p,q$ in $\curve$, as well as a basepoint $p_0\in\curve$ and an odd half period $z$, we define the prime form $E(p,q)$:
\beq
E(p,q):={\theta_{\vec{z}}(\vec{u}(p)-\vec{u}(q))\over \sqrt{dh_{\vec{z}}(p) dh_{\vec{z}}(q)}}
\eeq
where $dh_{\vec{z}}(p)$ is the holomorphic form:
\beq
dh_{\vec{z}}(p):= \sum_{i=1}^g \left.{\pa \theta_{\vec{z}}(\vec{u})\over \pa u_i}\right|_{\vec{u}=\vec{0}}\, du_i(p)
\eeq
\ed

\bt
The abelian differentials can be written:
\beq
dS_{q_1,q_2}(p) = d\ln{E(p,q_1)\over E(p,q_2)}
\eeq
\et

With these definitions, we have:
\beq
\L(p) = \gamma\,{\theta_{\vec{z}}(\vec{u}(p)-\vec{u}(\infty_-))\over \theta_{\vec{z}}(\vec{u}(p)-\vec{u}(\infty_+))}
\virg
\gamma:=\mathop{\rm lim}_{p\to\infty_+} \,{x(p)\,\theta_{\vec{z}}(\vec{u}(p)-\vec{u}(\infty_+))\over \theta_{\vec{z}}(\vec{u}(\infty_+)-\vec{u}(\infty_-))}
\eeq
\beq
H(p_\xi)={\theta_{\vec{z}}(\vec{u}(p_\xi)-\vec{u}(\infty_-))\theta_{\vec{z}}(\vec{u}(\infty_+)-\vec{u}(\pbar_\xi))\over \theta_{\vec{z}}(\vec{u}(p_\xi)-\vec{u}(\pbar_\xi))\theta_{\vec{z}}(\vec{u}(\infty_+)-\vec{u}(\infty_-))}
= -\gamma\,{\theta_{\vec{z}}(\vec{u}(\infty_+)-\vec{u}(\infty_-))\over \theta_{\vec{z}}(\vec{u}(p_\xi)-\vec{u}(\infty_+))^2}\,{dh_{\vec{z}}(p_\xi)\over dx(p_\xi)}
\eeq

\section{Asymptotics of orthogonal polynomials}

\subsection{One-cut case}

In the one-cut case, (i.e. genus zero algebraic curve), and if $V$ is a real potential, there is only one dominant saddle point if $\xi\notin [a,b]$,
and two conjugated dominant saddle points if $x\in[a,b]$.
More generaly, there is a saddle point corresponding to each determination of $p_\xi$ such that $x(p_\xi)=\xi$.
i.e. $p_\xi$ and $\pbar_\xi=1/p_\xi$.
The dominant saddle point is the one such that $\Re(V_{\rm eff}(p_\xi)-V(\xi))$ is minimal.
The two cols have a contribution of the same order if:
\beq
\Re V_{\rm eff}(p_\xi) = \Re V_{\rm eff}(\pbar_\xi)
\eeq
i.e. if $\xi$ is such that:
\beq\label{defcutsVeff}
\Re \int_{\pbar_\xi}^{p_\xi} W(x)dx =0
\eeq
If the potential is real, it is easy to see that the set of points which satisfy \ref{defcutsVeff} is $[a,b]$,
in general, it is a curve in the complex plane, going from $a$ to $b$, we call it the cut $[a,b]$ (similar curves were studied in \cite{moore}).

Then we have:
\begin{itemize}
\item For $x\notin[a,b]$, we write $\xi={a+b\over 2} + \gamma (p_\xi+1/p_\xi)$, $\gamma={b-a\over 4}$:
\beq
p_n(\xi) \sim \,\sqrt{H(p_\xi)}\,\left(\L(p_\xi)\right)^{n-N}\,\e^{-{N\over 2}(V_{\rm eff}(p_\xi)-V(\xi))} (1+O(1/N))
\eeq
i.e.
\beq
p_n(\xi) \sim \,\sqrt{\gamma\over x'(p_\xi)}\,\left(\gamma\, p_\xi\right)^{n-N}\,\e^{-{N\over 2}(V_{\rm eff}(p_\xi)-V(\xi))} (1+O(1/N))
\eeq

\item For $x\in[a,b]$, i.e. $p$ is on the unit circle $p=\e^{i\phi}$, $\xi={a+b\over 2}+2\gamma\cos\phi$:
\bea
p_n(\xi) &\sim& \,\sqrt{H(p_\xi)}\,\left(\L(p_\xi)\right)^{n-N}\,\e^{-{N\over 2}(V_{\rm eff}(p_\xi)-V(\xi))} (1+O(1/N)) \cr
&& + \,\sqrt{H(\pbar_\xi)}\,\left(\L(\pbar_\xi)\right)^{n-N}\,\e^{-{N\over 2}(V_{\rm eff}(\pbar_\xi)-V(\xi))} (1+O(1/N))
\eea
i.e.
\beq
p_n(\xi) \sim {\gamma^{n-N}\over \sqrt{2\sin\phi(\xi)}}\,2\cos{\left(N\pi\int_a^\xi \rho(x)dx - (n-N+{1\over 2}) \phi(\xi) + \alpha\right)} (1+O(1/N))
\eeq
i.e. we have an oscillatory behaviour

$$\begin{array}{r}
{\epsfxsize 10cm\epsffile{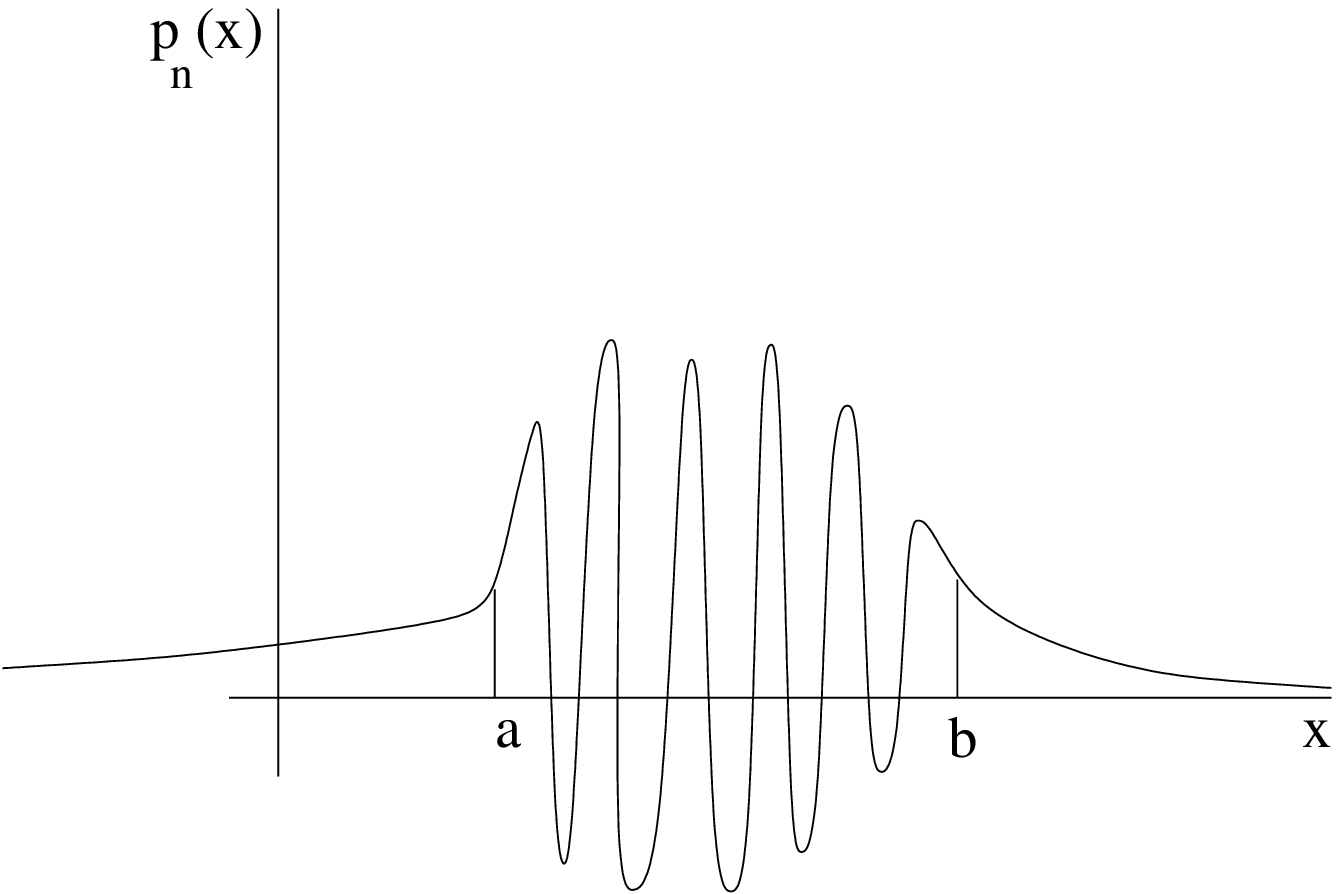}}
\end{array}$$

\end{itemize}

\subsection{Multi-cut case}
\label{sectmulticutasymp}

In the multicut case, in addition to having saddle-points corresponding to both determinantions of $p_\xi$,
we have a saddle point for each filling fraction configuration $n_1,\dots, n_s$ with $\sum_{i=1}^s n_i=n$.
We write:
\beq
\epsilon_i = {n_i\over N}
\eeq

The saddle point corresponding to filling fractions which differ by a  few units, contribute to the same order,
and thus cannot be neglected. One has to consider the sommation over filling fractions \cite{BDE}.

Thus, one has to consider the action of a saddle point as a function of the filling fractions.
We leave as an exercise for the reader to prove that the derivatives of $F$ are given by:
\beq
{\pa F\over \pa \epsilon_i} = -\oint_{{\cal B}_i} W(x)dx
\eeq
and:
\beq
{\pa^2 F\over \pa \epsilon_i\pa T} = -2i\pi (u_i(\infty_+)-u_i(\infty_-))
\eeq
\beq
{\pa^2 F\over \pa \epsilon_i\pa h} = -2i\pi (u_i(p_\xi)-u_i(\infty_+))
\eeq
\beq\label{dFtauij}
{\pa^2 F\over \pa \epsilon_i\pa \epsilon_j} = -2i\pi \tau_{ij}
\eeq
The last relation implies that $\Re F$ is a convex function of $\epsilon$, thus it has a unique minimum:
\beq
\vec\epsilon^*
\virg
\Re \left.{\pa F\over \pa \epsilon_i}\right|_{\vec\epsilon=\vec\epsilon^*} = 0
\eeq
We write:
\beq
\zeta_i := -{1\over 2i\pi} \left.{\pa F\over \pa \epsilon_i}\right|_{\vec\epsilon=\vec\epsilon^*}
\virg \zeta_i\in \R
\eeq

We thus have the Taylor expansion:
\bea
F(T,h,\vec\epsilon)
&\sim& F(1,0,\vec\epsilon^*) -2i\pi \vec\zeta^t (\vec\epsilon-\vec\epsilon^*)  +(T-1) {\pa F\over \pa T} + {h\over 2} (V_{\rm eff}(p_\xi)-V(\xi)) \cr
&& +{(T-1)^2\over 2} {\pa^2 F\over \pa T^2}-(T-1)h \ln{\L(p_\xi)}-{h^2\over 2} \ln{H(p_\xi)}\cr
&& -2i\pi (\vec\epsilon-\vec\epsilon^*)^t \tau (\vec\epsilon-\vec\epsilon^*)
-2i\pi (T-1) (\vec\epsilon-\vec\epsilon^*)^t  (\vec{u}(\infty_+)-\vec{u}(\infty_-)) \cr
&& -2i\pi h (\vec\epsilon-\vec\epsilon^*)^t (\vec{u}(p_\xi)-\vec{u}(\infty_+)) + \dots
\eea

Thus:
\bea
Z
&\sim& \sum_I C_I \e^{-N^2 F(\{x\}_I)} \cr
&\sim& \sum_{p=p_\xi,\pbar_\xi}
\e^{-N^2 F(1,0,\vec\epsilon^*)}\e^{N^2\left(-(T-1) {\pa F\over \pa T} - {h\over 2} (V_{\rm eff}(p)-V(\xi)) -{(T-1)^2\over 2} {\pa^2 F\over \pa T^2}+(T-1)h \ln{\L(p)}+{h^2\over 2} \ln{H(p)}\right)}\cr
&& \sum_{\vec{n}}
\e^{i\pi (\vec{n}-N\vec\epsilon^*)^t \tau (\vec{n}-N\vec\epsilon^*)}
\e^{2i\pi N \vec\zeta^t (\vec{n}-N\vec\epsilon^*)} \cr
&& \qquad \e^{2i\pi N(T-1) (\vec{n}-N\vec\epsilon^*)^t  (\vec{u}(\infty_+)-\vec{u}(\infty_-)) }
\e^{2i\pi Nh (\vec{n}-N\vec\epsilon^*)^t (\vec{u}(p)-\vec{u}(\infty_+))} \cr
\eea
In that last sum, because of convexity, only values of $\vec{n}$ which don't differ from $N\vec\epsilon^*$ form more than a few units, contribute substantialy.
Therefore, up to a non perturbative error (exponentialy small with $N$), one can extend the sum over the $n_i$'s to the whole $\Z^g$, and recognize a $\theta$-function (see \ref{deftheta}):
\bea
Z
&\sim& \sum_{p=p_\xi,\pbar_\xi}
\e^{-N^2 F(1,0,\vec\epsilon^*)}\e^{N^2\left((T-1) {\pa F\over \pa T} + {h\over 2} (V_{\rm eff}(p)-V(\xi)) +{(T-1)^2\over 2} {\pa^2 F\over \pa T^2}+(T-1)h \ln{\L(p)}+{h^2\over 2} \ln{H(p)}\right)}\cr
&& \e^{i\pi N^2 \vec\epsilon^{*t} \tau \vec\epsilon^*}
\e^{-2i\pi N^2 \vec\zeta^t \vec\epsilon^*}
\e^{-2i\pi N^2(T-1) \vec\epsilon^{*t}  (\vec{u}(\infty_+)-\vec{u}(\infty_-)) }
\e^{-2i\pi N^2 h \vec\epsilon^{*t} (\vec{u}(p)-\vec{u}(\infty_+))} \cr
&&
\theta(
N (\vec\zeta-\tau \vec\epsilon^*)
+N(T-1) (\vec{u}(\infty_+)-\vec{u}(\infty_-))
+Nh (\vec{u}(p)-\vec{u}(\infty_+))
,\tau) \cr
\eea
with $T-1={n-N\over N}$ and $h=0$ or $h=1/N$, we get the asymptotics:
\bea\label{asympmulticut}
p_n(\xi) &\sim& \sum_{x(p)=\xi}
\sqrt{H(p)}\, (\L(p))^{n-N}\,\e^{-{N\over 2} (V_{\rm eff}(p)-V(\xi))}\, \e^{-2i\pi N \vec\epsilon^{*t} (\vec{u}(p)-\vec{u}(\infty_+))} \cr
&&
{\theta(N (\vec\zeta-\tau \vec\epsilon^*) +(n-N) (\vec{u}(\infty_+)-\vec{u}(\infty_-)) +(\vec{u}(p)-\vec{u}(\infty_+)) ,\tau)
\over \theta(N (\vec\zeta-\tau \vec\epsilon^*) +(n-N) (\vec{u}(\infty_+)-\vec{u}(\infty_-))  ,\tau)}
 \cr
\eea

Again, depending on $\xi$, we have to choose the determination of $p_\xi$ which has the minimum energy.
If we are on a cut, i.e. if condition \ref{defcutsVeff} holds, both determinations contribute.
To summarize, outside the cuts, the sum \ref{asympmulticut} reduces to only one term, and along the cuts, the sum \ref{asympmulticut} contains two terms.

\section{Conclusion}

We have shown how the asymptotics of orthogonal polynomials (a notion related to integrability) is deeply related to algebraic geometry.
This calculation can easily be extended to many generalizations, for multi-matrix models \cite{eynchain, eynchaint, BEHAMS, eynhabilit}, non-hermitean matrices ($\beta=1,4$) \cite{eynbetapol}, rational potentials \cite{BEHsemiclas}, ...

\subsection*{Aknowledgements}
The author wants to thank the organizer of the Les Houches summer school Applications of Random Matrices in Physics
 June 6-25 2004.

\end{document}